\documentclass[twocolumn,pra,aps,showpacs,superscriptaddress]{revtex4-1}
\usepackage[english]{babel}
\usepackage[utf8]{inputenc}

\usepackage{url,psfrag,graphicx}
\usepackage{dcolumn}
\usepackage{amsmath,amssymb,amsthm}
\usepackage{bm}
\usepackage{pstricks}
\usepackage{hyperref}
\usepackage{epsfig,placeins}
\usepackage{mathtools}
\usepackage{commath}

\def\<{\left<}
\def\>{\right>}
\def\ket|#1>{\left|#1\right>}
\def\bra<#1|{\left<#1\right|}
\def\elem<#1|#2|#3>{\left<#1\right|#2\left|#3\right>}
\def\[{\left[}
\def\]{\right]}
\def\bJ{\textbf{J}}

\def\Tr{\text{\rm Tr}}

\def\({\left(}
\def\){\right)}

\def\R{{\mathbb R}}

\def\Re{\textrm{Re}}

\def\beq{\begin{equation}}
\def\eeq{\end{equation}}

\begin{document}

\title[Short Title]{Casimir forces on deformed fermionic chains}

\author{Begoña Mula}
\affiliation{Dto. Física Interdisciplinar, Universidad Nacional de
  Educación a Distancia (UNED), Madrid, Spain}

\author{Silvia N. Santalla}
\affiliation{Dto. Física \&\ GISC, Universidad Carlos III de Madrid,
  Leganés, Spain}

\author{Javier Rodríguez-Laguna}
\affiliation{Dto. Física Fundamental, Universidad Nacional de
  Educación a Distancia (UNED), Madrid, Spain}

\date{October 20, 2020}

\begin{abstract}
We characterize the Casimir forces for the Dirac vacuum on
free-fermionic chains with smoothly varying hopping amplitudes, which
correspond to (1+1)D curved spacetimes with a static metric in the
continuum limit. The first-order energy potential for an obstacle on
that lattice corresponds to the Newtonian potential associated to the
metric, while the finite-size corrections are described by a curved
extension of the conformal field theory predictions, including a
suitable boundary term. We show that, for weak deformations of the
Minkowski metric, Casimir forces measured by a local observer at the
boundary are metric-independent. We provide numerical evidence for our
results on a variety of (1+1)D deformations: Minkowski, Rindler,
anti-de Sitter (the so-called rainbow system) and sinusoidal metrics.
\end{abstract}

\maketitle


\section{Introduction}
\label{sec:intro}

The quantum vacuum on a static spacetime is nothing but the ground
state (GS) of a certain Hamiltonian. Therefore, it is subject to
quantum fluctuations which help minimize its energy. Yet, these
fluctuations are clamped near the boundaries, giving rise to the
celebrated {\em Casimir effect} \cite{Casimir.48}, see
\cite{Klimchitskaya.06} for experimental confirmations. Its relevance
extends away from the quantum realm, with applications to thermal
fluctuations in fluids \cite{Kardar.99}. Its initial description
required two infinite parallel plates, giving rise to an attractive
force between them. In fact, this attraction was rigorously proved for
identical plates by Kenneth and Klich \cite{Kenneth.06}, yet the force
can become repulsive or even cancel out when the boundary conditions
do not match \cite{Asorey.13}. The special features of fermionic 1D
systems have also been considered \cite{Sundberg.04,Zhabinskaya.08}.

For fields subject to conformal invariance, the Casimir force is
associated to the {\em conformal anomaly}, measured by the central
charge in 2D conformal field theory (CFT), $c$
\cite{Cardy.84,Cardy.86,DiFrancesco,Mussardo}. The expression for the
energy contains a non-universal contribution proportional to the
system size, plus finite-size corrections of order $O(1/N)$ which are
fixed by conformal invariance. Moreover, conformal invariance is
strong enough to yield an analytical expression for the Casimir forces
in presence of arbitrarily shaped boundaries \cite{Bimonte.13}.

The peculiarities of Casimir forces in curved spacetimes have been
considered by several authors \cite{DeWitt.75}. The problem is already
difficult for static spacetimes and weak gravitational fields
\cite{Sorge.05,Sorge.19,Nouri.10,Nazari.12}. The Casimir force takes
the same form on weak static gravitational fields at first-order, when
coordinate differences are substituted by actual distances, although
with non-trivial second-order corrections. Interestingly, the Casimir
effect has been put forward as a possible explanation of the
cosmological constant, making use of Lifshitz theory
\cite{Leonhardt.19,Leonhardt.20}.

Even if our technological abilities do not allow us to access direct
measurements of the Casimir effect in curved spacetimes, we are aware
of possible strategies to develop quantum simulators using current
technologies, such as ultracold atoms in optical lattices
\cite{Lewenstein.12}. Concretely, it has been shown that the Dirac
vacuum on certain static spacetimes can be characterized in such a
quantum simulator \cite{Boada.11}, and an application has been devised
to measure the Unruh radiation, including its non-trivial dimensional
dependence \cite{Laguna_unruh.17,Takagi.86,Louko.18}. The key insight
is the use of {\em curved optical lattices}, in which fermionic atoms
are distributed on a flat optical lattice with inhomogeneous hopping
amplitudes, thus simulating a position-dependence index of refraction
or, in other terms, an {\em optical metric}.

Dirac vacua in such curved optical lattices present quite novel
properties. When the background metric is negatively curved, i.e.:
(1+1)D anti-de Sitter (AdS), the entanglement entropy (EE) may violate
maximally the area law \cite{Eisert.10}, forming the so-called {\em
  rainbow state} \cite{Vitagliano.10,Ramirez.14,Ramirez.15}.
Interestingly, the EE of blocks within the GS of a (1+1)D system with
conformal invariance is fixed by CFT
\cite{Holzhey.94,Vidal.03,Calabrese.04,Calabrese.09}. Such conformal
arguments can be extended to a statically deformed (1+1)D system, and
the EE of the rainbow system was successfully predicted
\cite{Laguna.17}, along with other interesting magnitudes, such as the
entanglement spectrum, entanglement contour and entanglement
Hamiltonian \cite{Tonni.18,MacCormack.18}.

The aim of this article is to extend the aforementioned (1+1)D CFT
predictions on curved backgrounds to characterize the Casimir force
for the fermionic vacuum on curved optical lattices. This article is
organized as follows. In Sec. \ref{sec:model} we describe our physical
system and summarize the CFT techniques employed to evaluate the EE on
curved backgrounds, providing some examples. Sec. \ref{sec:forces}
characterizes the Casimir forces on curved optical lattices, using the
same example backgrounds, emphasizing the role of universality in the
finite-size corrections. The article closes with a series of
conclusions and proposals for further work.


\section{Fermions on curved optical lattices}
\label{sec:model}

\begin{figure}
  \includegraphics[width=8.5cm]{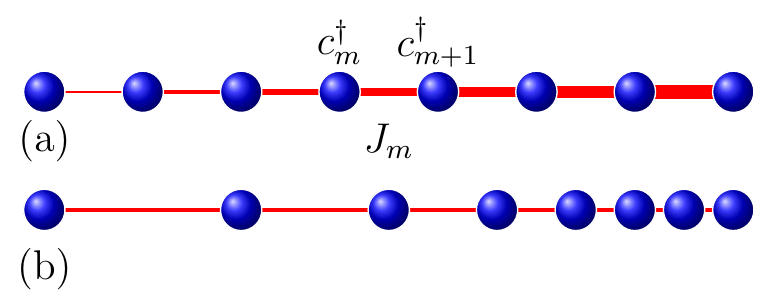}
  \caption{(a) Illustration of an inhomogeneous chain with $N=8$
    sites. (b) Corresponding positions after the deformed coordinates
    $\tilde x$.}
  \label{fig:illust}
\end{figure}

Let us consider an open fermionic chain with (even) $N$ sites, whose
Hilbert space is spanned by creation operators $c^\dagger_m$, $m\in
\{1,\cdots,N\}$ following standard anticommutation relations. We can
define an inhomogeneous hopping Hamiltonian,

\beq
H(\bJ)_N=-\sum_{m=1}^{N-1} J_m c^\dagger_m c_{m+1} + \text{h.c.},
\label{eq:ham}
\eeq
where $\bJ=\{J_m\}_{m=1}^{N-1}$ are the {\em hopping amplitudes},
$J_m\in \R^+$ referring to the link between sites $m$ and $m+1$, see
Fig. \ref{fig:illust} (a). In order to obtain some physical intuition,
let us remember that the set of $\{J_m\}$ constitutes a
position-dependent Fermi velocity, i.e.: a signal takes a time of
order $J_m^{-1}$ to travel between sites $m$ and $m+1$. If the
$\{J_m\}$ are {\em smooth enough}, we can assume $J_m=J(x_m)$ for a
certain smooth function $J(x)$, with $x_m=m\Delta x$. Unless otherwise
specified, we will use $\Delta x=1$.

It can be proved that Eq. \eqref{eq:ham} is a discretized version of
the Hamiltonian for a Dirac fermion on a curved (1+1)D spacetime with
a static metric of the form \cite{Boada.11,Ramirez.15,Laguna.17}

\beq
ds^2 = -J^2(x) dt^2 + dx^2,
\label{eq:metric}
\eeq
i.e. a spacetime metric with a position dependent speed of light or,
equivalently, a modulated {\em index of refraction}. Defining $\tilde
x(x)$ such that

\beq
d\tilde x=\frac{dx}{J(x)},
\label{eq:defxtilde}
\eeq
we have

\beq
ds^2=J^2(x)(-dt^2+d\tilde x^2),
\label{eq:altmetric}
\eeq
which is conformally equivalent to the Minkowski metric. This
deformation is illustrated in Fig. \ref{fig:illust} (b): sites get
closer when the $J_m$ associated to their link is large, giving rise
to an homogeneous effective hopping amplitude.

Conformal equivalence between metrics \eqref{eq:altmetric} and the
Minkowski metric suggests that conformal field theory (CFT) techniques
might describe the universal properties of low-energy eigenstates of
Hamiltonian \eqref{eq:ham}. Indeed, we will show that this is the
case, once those universal properties have been ascertained.

Some interesting metrics fall into this category. If $J(x)=J_0$ is a
constant, we recover Minkowski spacetime on a finite spatial
interval. The Rindler metric, which is the spacetime structure
perceived by an observer moving with constant acceleration $a$ in a
Minkowski metric, is described by

\begin{equation}
  J(x)=J_0+ax.
  \label{eq:rindler_J}
\end{equation}
Notice that it presents an {\em horizon} at $x_h=-J_0/a$, where the
local speed of light vanishes. Information can not cross this point,
thus separating spacetime into two {\em Rindler wedges}
\cite{Wald}. We will consider some other choices for the hopping
amplitudes, such as the {\em sine metric},

\begin{equation}
  J(x)=J_0+A\sin\(k x\),
  \label{eq:sine_J}
\end{equation}
or a rainbow metric given by

\begin{equation}
  J(x)=J_0\exp\(-h\left|x-\frac{N}{2}\right|\),
  \label{eq:rainbow_J}
\end{equation}
for $h\geq 0$, with $h=0$ corresponding to the Minkowski case. This
metric has constant negative curvature except at the center, $x=N/2$,
thus resembling an anti-de Sitter (adS) space, and has been considered
recently because its vacuum presents volumetric entanglement
\cite{Vitagliano.10,Ramirez.14,Ramirez.15,Laguna.17,Tonni.18,MacCormack.18}.
Unless otherwise stated, we will always assume $J_0=1$.

\subsection{Free fermions on the lattice}

The exact diagonalization of Hamiltonian \eqref{eq:ham} is a
straightforward procedure which only involves the solution of the
associated single-body problem. Let us define the hopping matrix,
$T_{ij}=T_{ji}=-J_i\delta_{i,j+1}$, such that

\beq
H(\bJ)_N=-\sum_{i,j} T_{ij} c^\dagger_i c_j,
\label{eq:discrH}
\eeq
then we can diagonlize the hopping matrix, $T_{ij}=\sum_k U_{i,k}
\epsilon_k \bar U_{j,k}$, where $\epsilon_k$ are the single-body
energies and the columns of $U_{i,k}$ represent the single-body
modes. The GS of Hamiltonian \eqref{eq:ham} can be written as
$\ket|\Psi>=\prod_{k=1}^{N/2} b^\dagger_k \ket|0>$, where $\ket|0>$ is
the Fock vacuum and $b^\dagger_k = \sum_i U_{i,k} c^\dagger_i$.

The system presents particle-hole symmetry,
$\epsilon_k=-\epsilon_{N+1-k}$, with $U_{i,k}=(-1)^i U_{i,N+1-k}$. At
half-filling the local density is always homogeneous,
$\<c^\dagger_nc_n\>=1/2$ for all $n$, independently of the metric. For
the Minkowski metric,

\beq
\<c^\dagger_n c_{n+1}\>= \sum_{k=1}^{N/2} \bar U_{n,k}
U_{n+1,k}\approx {c_0 \over 2} \equiv {1\over\pi},
\label{eq:localcorr}
\eeq
plus a correction term presenting parity oscillations, related to the
fact that the Fermi momentum is $k_F=\pi/2$.


\subsection{CFT and entanglement for curved lattice fermions}
\label{sec:entropy}

Let us provide a cursory summary of the application of CFT techniques
to the characterization of the entanglement structure of the fermionic
vacuum on curved optical lattices.

The von Neumann entanglement entropy (EE) of a block $A$ of a pure
state $\ket|\Psi>$ is defined as

\begin{equation}
  S_A=-\Tr\[\rho\log\rho_A\],
  \label{eq:vonneumann}
\end{equation}
where $\rho_A=\Tr_{\bar A} \ket|\Psi>\bra<\Psi|$ is the reduced density
matrix for block $A$. In the case of Gaussian states, which follow
Wick's theorem, this magnitude can be determined from the two-point
correlation function with low computational effort
\cite{Peschel.03}. Following \cite{Calabrese.04,Calabrese.09}, the EE
of a lateral block $A=\{1,\cdots,\ell\}$ of the GS of a conformal system
with central charge $c$ on a chain with $N$ sites can be written as

\begin{equation}
  S(\ell)={c\over 6}\log\( {N\over\pi\Delta x}
  \sin\( {\pi \ell \over N} \) \) + S_{\text{non-univ}}.
    \label{eq:S_cft}
\end{equation}
where $c=1$ for free fermions, $\Delta x$ is the UV cutoff and
$S_{\text{non-univ}}$ is a non-universal contribution containing a
constant term and parity oscillations which has been explicitly
computed for the free-fermionic case \cite{Jin.04,Fagotti.11}.

Expression \eqref{eq:S_cft} has been successfully extended to evaluate
entanglement entropies on the GS of Hamiltonian \eqref{eq:ham}
\cite{Laguna.17,Tonni.18}. When Dirac fermions are inserted in a
smooth static optical metric of the type \eqref{eq:metric}, the EE
deforms appropriately, i.e. the block lengths must be transformed via
Eq. \eqref{eq:defxtilde},

\beq
\ell\to\tilde\ell=\tilde x(\ell\Delta x)=\int_{x_0}^{\ell\Delta x}
                         {dx\over J(x)}
\approx \sum_{p=1}^{\ell-1} {\Delta x\over J_p},
\label{eq:elltilde}
\eeq
while $\tilde N=\tilde x(N\Delta x)$. We must also take into account
the transformation of the UV cutoff,

\beq
\Delta x \;\to\; \Delta\tilde x(\ell)=\frac{\Delta x}{J(\ell)}.
\label{eq:uvdef}
\eeq
Thus, we obtain

\begin{equation}
  S(\ell)={c\over 6}\log\( {\tilde N\over\pi\Delta\tilde x} \sin\(
  {\pi\tilde\ell\over\tilde N} \)\) + S_{\text{non-univ}}.
  \label{eq:EE_deformed}
\end{equation}

Concretely, in \cite{Laguna.17,Tonni.18} the EE for lateral blocks
within the GS of the rainbow Hamiltonian \eqref{eq:ham} using
\eqref{eq:rainbow_J} was obtained using

\begin{align}
  \Delta\tilde x & =e^{-h|N/2-\ell|}\Delta x,\\
  h \tilde N & =2(e^{hN/2}-1)\Delta x, \\
  h \tilde\ell & =
  \begin{cases}
    \(e^{hN/2}-e^{h(N/2-\ell)}\) \Delta x, &
    \text{if } \ell\leq  N/2, \\
    \(e^{hN/2}+e^{h(\ell-N/2)}\) \Delta x, &
    \text{if } \ell\geq N/2.\\
  \end{cases}
  \label{eq:EE_rainbow}
\end{align}

In the limit $h\ell\gg 1$, the EE of a block of size $\ell\leq N/2$ becomes

\beq
S(\ell)\approx {ch\over 6} \ell + S_{\text{non-univ}},
\eeq
i.e. it yields a volume law for entanglement \cite{Tonni.18},
violating maximally the so-called area law of entanglement
\cite{Eisert.10}. We can also apply Eq. \eqref{eq:EE_deformed} to the
case of the Rindler metric, where we find

\begin{equation}
  S(\ell)={1\over 6}\log\({\ell\log N\over \pi\Delta x} \sin\({\pi \log(N/\ell)
    \over \log N }\) \) + S_{\text{non-univ}}.
  \label{eq:EE_rindler}
\end{equation}

\begin{figure}
  \includegraphics[width=9cm]{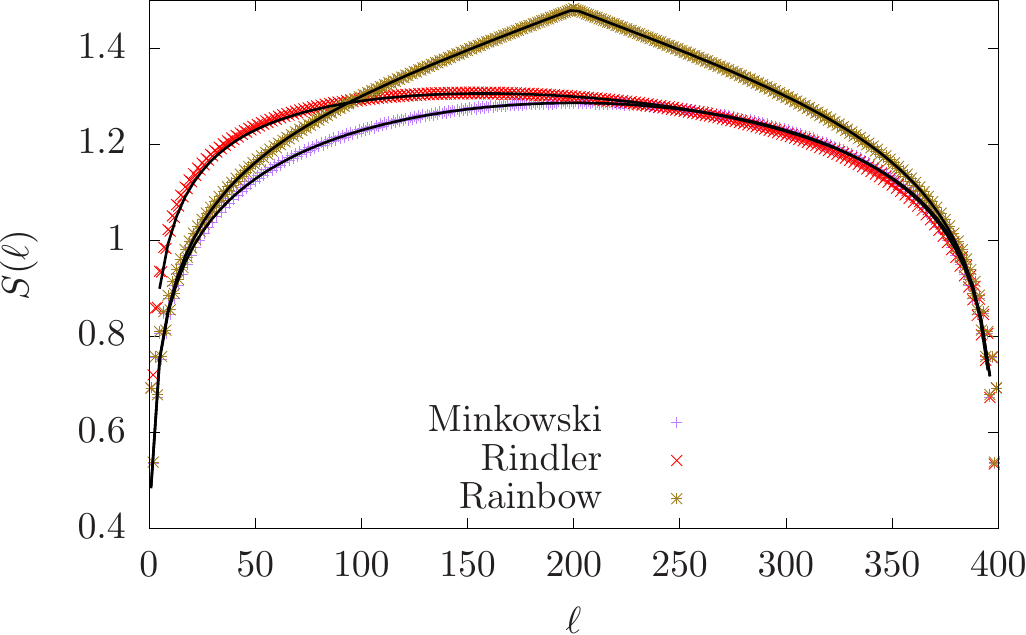}
  \caption{Entanglement entropy of the GS of free fermionic systems on
    a optical chain with $N=400$ for three different metrics:
    Minkowski, rainbow (Eq. \eqref{eq:rainbow_J} with $h=0.01$) and
    Rindler (Eq. \eqref{eq:rindler_J} with $a=2$), using the
    procedures of \cite{Peschel.03} The continuous lines are the CFT
    prediction, given by Eq. \eqref{eq:EE_deformed}, with a
    non-universal constant term added.}
  \label{fig:ee}
\end{figure}

The validity of these expressions can be checked in Fig. \ref{fig:ee},
where we have plotted the entropy $S_A$ as a function of the block
size $l$ for three systems using $N=400$: the Minkowski case,
Eq. \eqref{eq:S_cft}, the rainbow case with $h=0.01$,
Eq. \eqref{eq:EE_deformed} with \eqref{eq:EE_rainbow}, and the Rindler
case with $a=2$, via Eq. \eqref{eq:EE_rindler}. Indeed, the
non-universal terms are present, which also carry parity oscillations,
but they are a small correction to the entanglement entropy as
predicted by the CFT.

The accuracy of the CFT prediction allows us to conjecture that free
Dirac fermions on curved optical lattices can be characterized by a
suitable deformation of a conformal field theory, expecting that the
non-universal terms will be small enough. We will put this conjecture
to the test in the next section.


\section{Casimir forces on curved optical lattices}
\label{sec:forces}

Let us characterize the Casimir forces on curved optical lattices in
successive approximations. First of all, we will show that the GS
energy of Hamiltonian \eqref{eq:ham} is proportional to the sum of the
hoppings in first-order perturbation theory. This will lead us to show
that the force felt by a classical obstacle immerse in that state will
be similar to the Newtonian gravitational force in the corresponding
metric. Then, we will reach the main result of this work: the
finite-size corrections to the Casimir energy are universal, and the
corresponding expressions are a deformed variant of the general CFT
form.

\subsection{Potential energy and correlator rigidity}

Let us consider a free fermionic chain of $N$ sites on a deformed
metric, following Eq. \eqref{eq:ham}. The exact vacuum energy can be
written as

\beq
E_N= - 2\sum_{p=1}^{N-1} J_p \;\Re\langle c^\dagger_p c_{p+1}\rangle.
\label{eq:energy}
\eeq
We can estimate this expression via perturbation theory, if we assume
that $J_p=J_0+\delta J_p$ and make use of
Eq. \eqref{eq:localcorr}. The result at first-order is

\beq
E_0 \approx - c_0 S_N,\qquad\text{where } S_N\equiv \sum_{p=1}^{N-1} J_p.
\eeq

\begin{figure}
  \includegraphics[width=8cm]{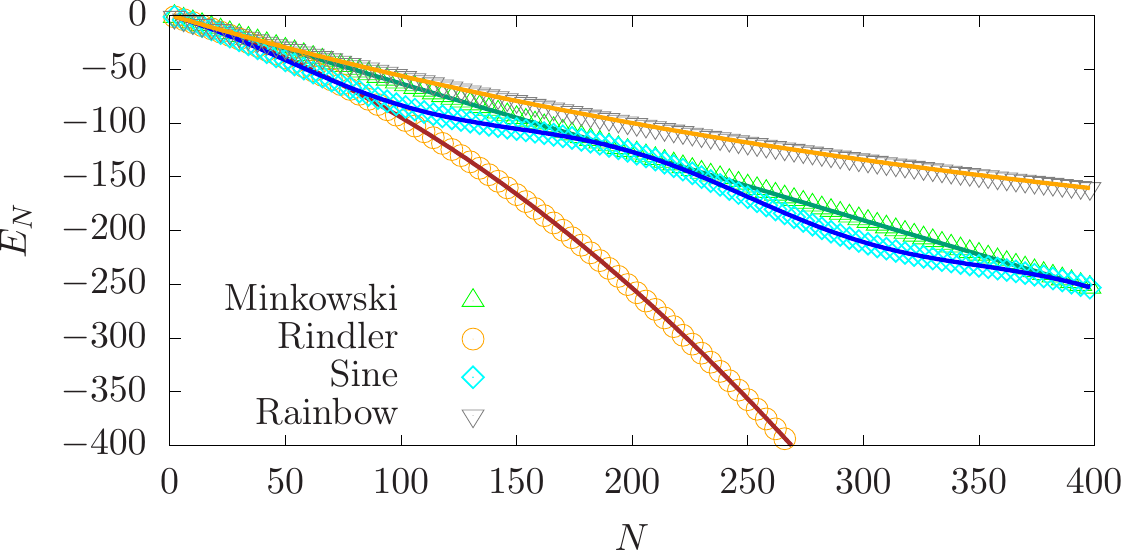}
  \includegraphics[width=8cm]{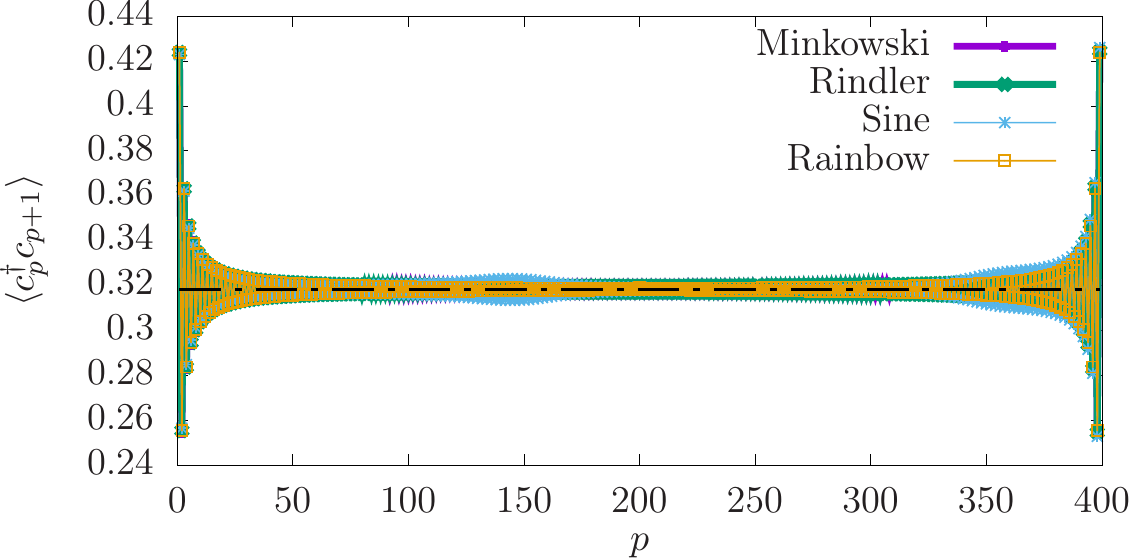}
  \caption{Top: Check of the bulk prediction for the energy,
    $E_0\approx - c_0S_N$ for four metrics: Minkowski, Rindler
    ($a=0.01$), Sine ($A=0.5$, $k=\pi/100$) and Rainbow ($h=5\cdot
    10^{-3}$. Numerical values are given in dots, while the
    theoretical prediction is provided in the full line. Bottom:
    Illustration of the correlator rigidity. Local correlators,
    $\langle c^\dagger_pc_{p+1}\rangle$ as a function of the position
    $p$ for the same four metrics.}
  \label{fig:corr}
\end{figure}

The validity of this approximation can be checked in the top panel of
Fig. \ref{fig:corr}, for four different metrics: Minkowski, Rindler,
Sine and Rainbow. The accuracy of our conjecture suggests that the
local correlators in the deformed vacuum are still homogeneous. In
fact, we will make the further claim that the {\em local correlators
  are rigid}, i.e. $\langle c^\dagger_p c_{p+1} \rangle \approx c_0/2$
for a weakly deformed metric. This claim has been checked
independently in the bottom panel of Fig. \ref{fig:corr}, where the
local correlators are shown for different deformations. Indeed, their
average values are still very close to $c_0=2/\pi$, and the only
substantial deviation is provided by the expected parity oscillations
which are well known in the Minkowski case.

A heuristic argument to understand correlator rigidity may be as
follows. For fermionic fields in Minkowski spacetime we have
$\<\psi(x)\psi(x+\Delta x)\> \sim \Delta x^{-1}$. After a deformation,
$\Delta x \to \Delta \tilde x = \Delta x/J(x)$. Yet, the fields
transform also as $\tilde\psi(x)=J^{1/2}(x)\psi(x)$, and the local
correlator remains invariant.

\bigskip

\begin{figure}
  \includegraphics[width=8cm]{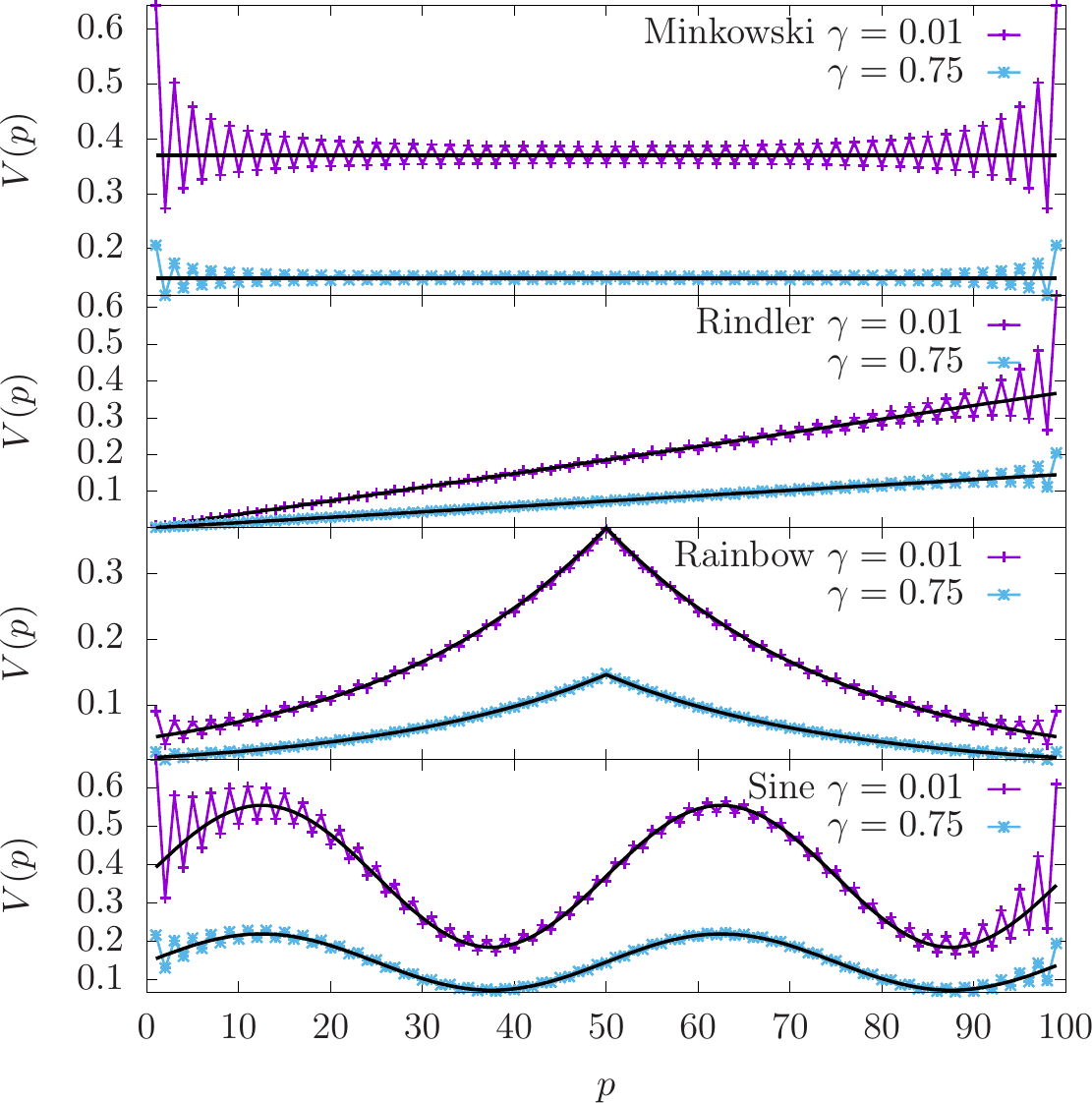}
  \caption{Potential energy $V(p)$ obtained by inhibiting the $p$-th
    hopping by a factor $\gamma$, $J_p\to \gamma J_p$, for four
    different metrics: Minkowski, Rindler ($a=0.01$), rainbow
    ($h=0.04$) and sinusoidal ($A=0.5$ and $k=2\pi/50$), always using
    $N=100$ and two values of $\gamma=0.01$ and $0.75$. In continuous
    line, we plot $J(x)$ mutiplied by a factor which only depends on
    $\gamma$.}
  \label{fig:potential}
\end{figure}

Let us consider a classical particle standing between sites $p$ and
$p+1$, which acts like an obstacle inhibiting the local hopping by a
factor $\gamma<1$, $J_p\to \gamma J_p$. Let us now evaluate the excess
energy of the deformed GS as a function of $p$, $V(p)=E_0(p)-E_0$,
which acts as a {\em potential energy function} for the obstacle. The
results are shown in Fig. \ref{fig:potential}, where we plot $V(p)$
for the same four different situations, using $N=100$ and both
$\gamma=0.01$ and $\gamma=0.75$. As $\gamma$ approaches 1 the trivial
case is recovered, i.e. the potential energy is equivalent to $E_0$.

The first salient feature of Fig. \ref{fig:potential} is that the
potential energy $V(p)$ resembles the hopping function $J(x)$, with
some strong parity oscillations. We are thus led to conjecture that a
classical particle moving on a static metric in (1+1)D would be
dragged by a force similar to the graviational pull. Making use of
Hellmann-Feynman's theorem, we see that

\beq
V(p) \approx - 2 J_p \Re \<c^\dagger_p c_{p+1}\>\approx -2J_p c_0.
\label{eq:potential}
\eeq

\subsection{Finite-Size Corrections}

The GS of a finite open chain of $N$ sites in Minkowski spacetime is
given by Cardy's expression
\cite{Cardy.84,Cardy.86,DiFrancesco,Mussardo}

\beq
E_N=-c_0 (N-1) - c_B - \dfrac{c\pi v_F}{24N} + O(N^{-2}),
\label{eq:cft}
\eeq
where $c$ is the associated central charge, $v_F$ is the Fermi
velocity and $c_0$ and $c_B$ are non-universal constants, which
correspond to the bulk energy per link and the boundary energy. Notice
that the last term is {\em universal}, since its form is fixed by
conformal invariance \cite{Cardy.84,Cardy.86,DiFrancesco,Mussardo},
but the bulk and boundary terms are not. The GS energy of Hamiltonian
\eqref{eq:ham} with $J_n=1$ follows Eq. \eqref{eq:cft} very
accurately, using $c=1$ for Dirac fermions, $v_F=2$, $c_0=2/\pi$ and
$c_B=4/\pi-1$.

Our main target is to generalize expression \eqref{eq:cft} to the case
of deformed backgrounds. Indeed, we may follow the guidelines of
Sec. \ref{sec:entropy} and attempt a substitution $x\to \tilde x$,
such that $d\tilde x/dx=J(x)^{-1}$, but {\em it will not work} for the
bulk and boundary terms. In that case, the bulk energy would become
proportional to $\tilde N$. Thus, in the rainbow case we should obtain
an energy term which grows exponentially with $N$ for any fixed $h>0$,
which is not found. Indeed, as we will show, that transformation is
only relevant for the universal term.

\medskip

Let us propose an extension of Eq. \eqref{eq:cft} to curved
backgrounds based on physical arguments, term by term.

\begin{itemize}
  \item The term $c_0 (N-1)$ stands for the bulk energy, which
    should be replaced by $c_0 S_N$, i.e. the sum of the $N-1$ first
    hopping amplitudes, multiplied by the local correlator term.
  \item The boundary term, $c_B$ should be proportional to the
    terminal hoppings, thus generalizing to $c_B(J_1+J_{N-1})/2$.
  \item The conformal correction is universal. Thus, it must be
    naturally deformed, changing $N^{-1}$ into $\tilde N^{-1}$, where
    $\tilde N$ is the effective length in deformed coordinates, given
    by $\tilde N=\sum_{i=1}^{N-1} J_i^{-1}$ (we let $\Delta x=1$).
\end{itemize}

Thus, we claim that the correct generalization of Eq. \eqref{eq:cft}
to curved optical lattices is given by

\beq
E_N=-c_0 S_N - {c_B\over 2}\(J_1+J_{N-1}\)
- \dfrac{c\pi v_F}{24\tilde N} + O(N^{-2}).
\label{eq:cft_curved}
\eeq

This expression can be more rigorously justified through a careful
analysis of the conformal field theory origin of Eq. \eqref{eq:cft},
and this is discussed in Appendix \ref{appendix1}.

The inverse of the deformed length $\tilde N^{-1}$ can be given an
interesting physical interpretation. Indeed, it is easy to recognize
$(N-1)\tilde N^{-1}$ as the {\em harmonic average} of the local speeds
of light, which can be understood as an {\em effective} Fermi
velocity, $\bar v_F$. Yet, for small deformations, the harmonic
average is similar (and lower than) the arithmetical average. Thus,
for the sake of simplicity, we approximate $\bar v_F \approx
2S_N/(N-1)$. Thus, we may provide an approximate version of
Eq. \eqref{eq:cft_curved} for a weakly deformed (1+1)D lattice,

\begin{equation}
  E_N \approx - c_0 S_N - {c_B\over 2} (J_1+J_{N-1})
  - {\pi S_N \over 12 N^2}.
  \label{eq:deformed}
\end{equation}

\subsection{Universality of Casimir forces in curved backgrounds}

Numerical checks of Eqs. \eqref{eq:cft_curved} or \eqref{eq:deformed}
must be subtle, because the finite-size correction is typically much
smaller than the bulk energy term. Let us consider an alternative
observable: the Casimir force measured by a local observer located at
the boundary. Since energy is associated to a frequency, local
energy measurements at site $x$ will be given by

\beq
E(x) = {E_N \over g_{00}^{1/2}(x)} = {E_N \over J_N}.
\label{eq:energy_observer}
\eeq

Such an observer will measure a force given by the covariant spatial
derivative of $F = -D_x E(x)$, taking the lattice spacing $\Delta x=1$
(see Appendix \ref{appendix2} for details) and changing the sign for
convenience, we define

\begin{equation}
  F_N \equiv {E_N-E_{N-2}\over J_{N-1}+J_{N-2}}.
  \label{eq:FN}
\end{equation}

Assuming smoothly varying hopping amplitudes we obtain

\begin{equation}
  F_N\equiv -c_0 - {c_B\over 2} \({J'_N\over J_N}\) -
  {\pi \over 12 N^2} + {\pi S_N \over 6 J_N N^3}.
  \label{eq:realforce}
\end{equation}
Let us consider the terms individually. The first, $c_0=2/\pi$, is
simply associated to the bulk energy. The second is a {\em boundary
  force}, which is absent from the homogeneous case, and will take a
leading role in some cases. For very weak deformations, $J_N\approx
J_0 + \delta J_N$ is a small deformation, we can assume that $S_N
\approx NJ_N$, and we obtain

\begin{equation}
  F_N \approx -c_0 -{c_B\over 2} \({J'_N\over J_N}\) + {\pi
    \over 12N^2}.
  \label{eq:force}
\end{equation}

Thus, we are led to the following claim: {\em Casimir forces on a
  weakly curved background are metric-independent when measured by a
  local observer at the boundary.} Indeed, consider an observer on a
classical obstacle located at site $p$. It will be subject both to a
left and a right Casimir forces. The bulk and boundary parts will
cancel out, and only the universal finite-size correction will
survive, yielding

\beq
F(p)=F_{N-p}-F_p={\pi\over 12}\({1\over (N-p)^2}-{1\over p^2}\).
\eeq

\medskip

\begin{figure}
  \includegraphics[width=8cm]{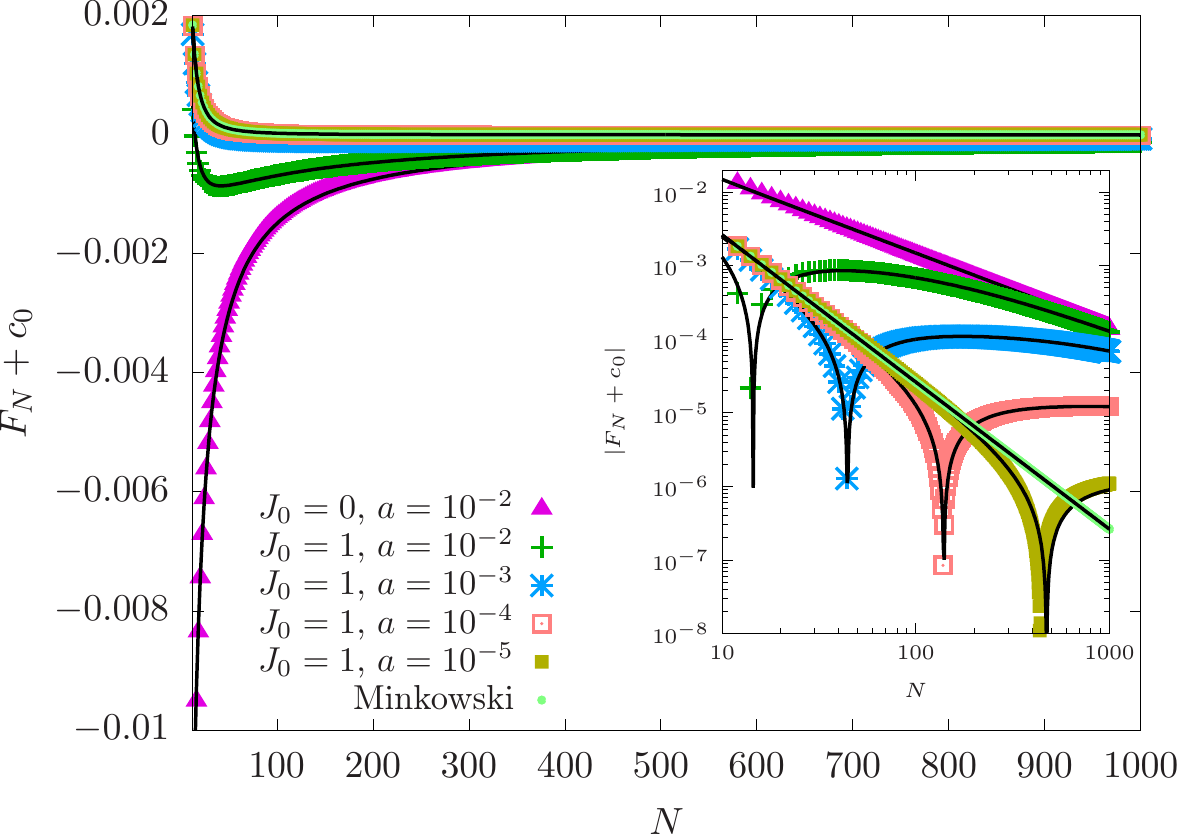}
  \includegraphics[width=8cm]{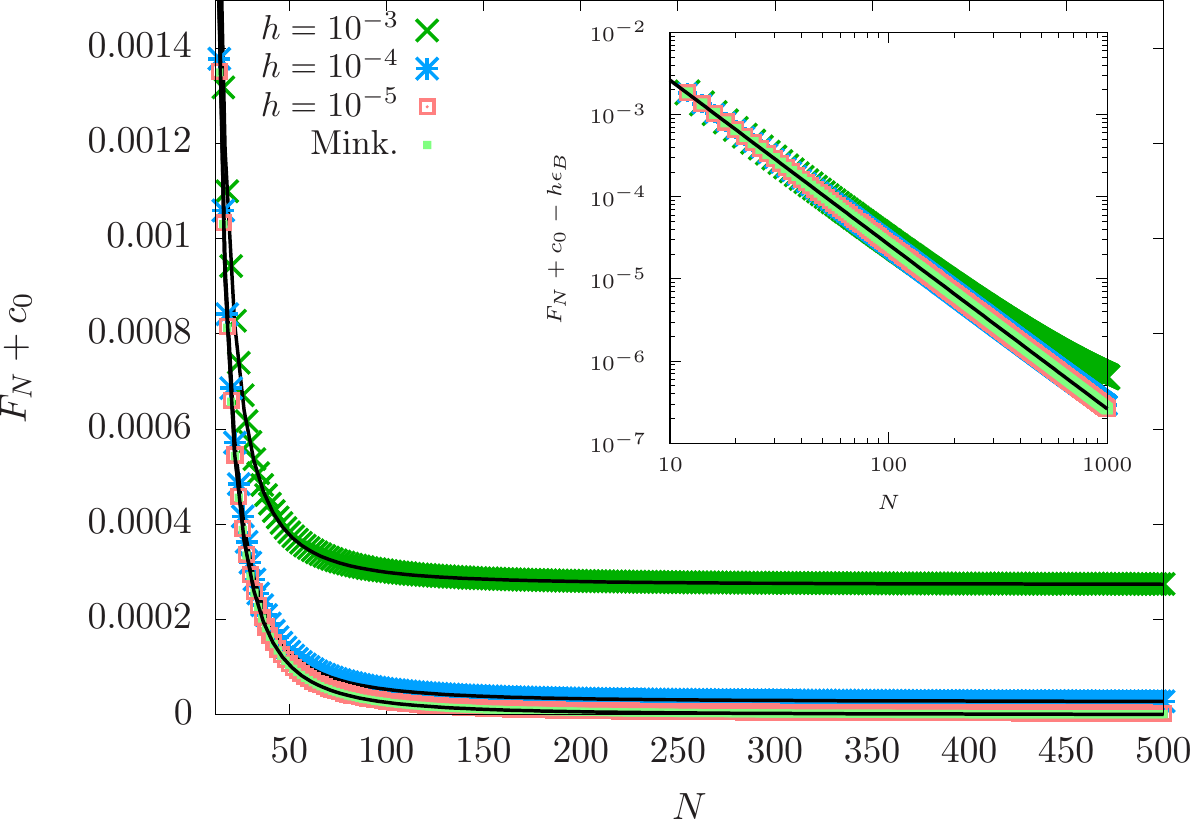}
  \includegraphics[width=8cm]{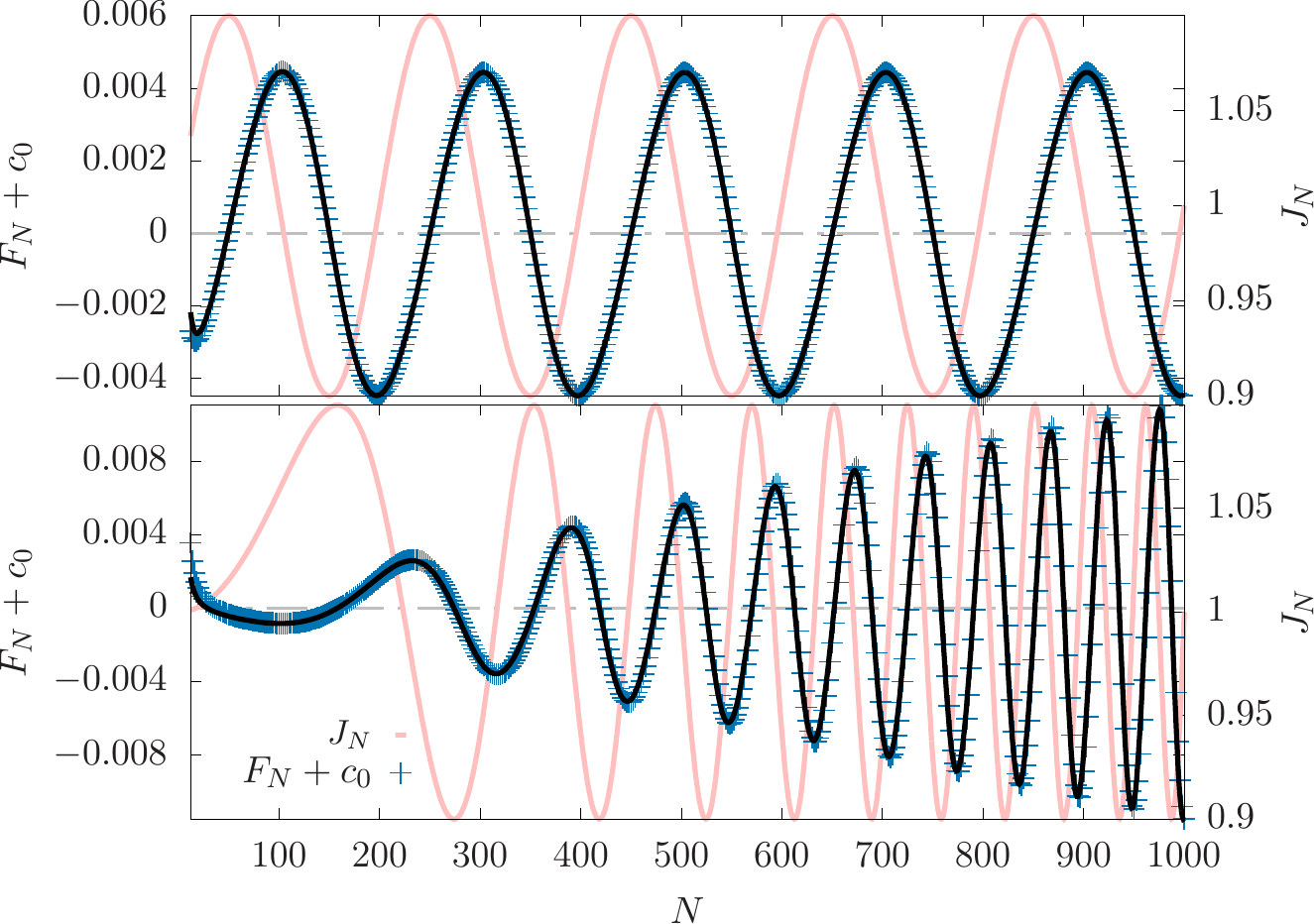}
  \caption{Casimir forces, $F_N+c_0$, for different metrics. Top:
    Rindler metric. Inset, log-log plot of $|F_N+c_0|$ as a function
    of $N$, in log-log scale. Notice most small systems are dominated
    by the CFT correction, while for larger sizes the boundary term
    $N^{-1}$ dominates. Center: Rainbow metric, we observe that
    $F+c_0$ tends to $\epsilon_Bh$. Inset: log-log plot of
    $F_N+c_0-\epsilon_Bh$. Bottom: Sinusoidal metric (top) and
    modulated frequency metric (bottom).}
  \label{fig:forces}
\end{figure}

The validity of expression \eqref{eq:force} can be checked in
Fig. \ref{fig:forces}. In all cases, the black continuous line is the
theoretical prediction, Eq. \eqref{eq:force}. The top panel shows the
forces $F_N+c_0$ as a function of $N$ for Rindler metrics of different
sizes, varying both $J_0$ and the acceleration $a$. We have included
the Minkowski case, which corresponds to $J_0=1$ and $a=0$, as one of
the limits. We notice that $F_N+c_0$ can be both positive and
negative, depending on the values of $J_0$ and the acceleration
$a$. This behavior is explained through our expression
\eqref{eq:force}: the boundary term scales like $N^{-1}$ and it is
always negative. Meanwhile, the universal conformal term scales like
$N^{-2}$ and is always positive. Thus, the prevalence of one or the
other explains the global behavior, but for large enough $N$ the
boundary term is always dominant. This trade-off can be visualized in
the inset, where we plot the absolute value $|F_N+c_0|$ as a function
of $N$ in log-log scale. For Minkowski, $J_0=1$ and $a=0$, the $1/N^2$
behavior extends for all sizes, but as soon as $a>0$ we observe a
small-$N$ behavior like $N^{-2}$ which performs a crossover into the
dominant $N^{-1}$ term beyond a finite size which scales as
$(J_0/a)^{1/2}$.

The central panel of Fig. \ref{fig:forces} shows the case of the
Casimir forces in the rainbow state, for which the boundary term is
constant: $J'_N/J_N=-h$ for all $N$. Thus, the behavior of $F_N+c_0$
corresponds merely to the CFT term, Eq. \eqref{eq:cft} with a constant
additive correction. This behavior is further clarified when this
constant is removed, and we observe the nearly perfect collapse of all
the forces in the inset of Fig. \ref{fig:forces} (center).

We have also considered is the sinusoidal metric,
Eq. \eqref{eq:sine_J}, where the boundary term dominates the force
for large $N$, while the CFT term dominates for low $N$, as we can see
in the bottom panel of Fig. \ref{fig:forces}. There, we can observe
the behavior of the hoppings (in pale pink), along with the forces and
their fit to expression \eqref{eq:force}. Indeed, the force
behaves like the derivative of the hopping function. In order to
highlight this behavior, we have considered yet another metric, given
by

\begin{equation}
  J_N = 1+A\sin(kN^2),
\end{equation}
i.e. a modulated frequency sinusoidal. The results are shown in the
bottom panel of Fig. \ref{fig:forces}, showing again an excellent
agreement between the theory and the numerical experiments.


\section{Casimir force in the inhomogeneous Heisenberg model}

\begin{figure}
  \includegraphics[width=8cm]{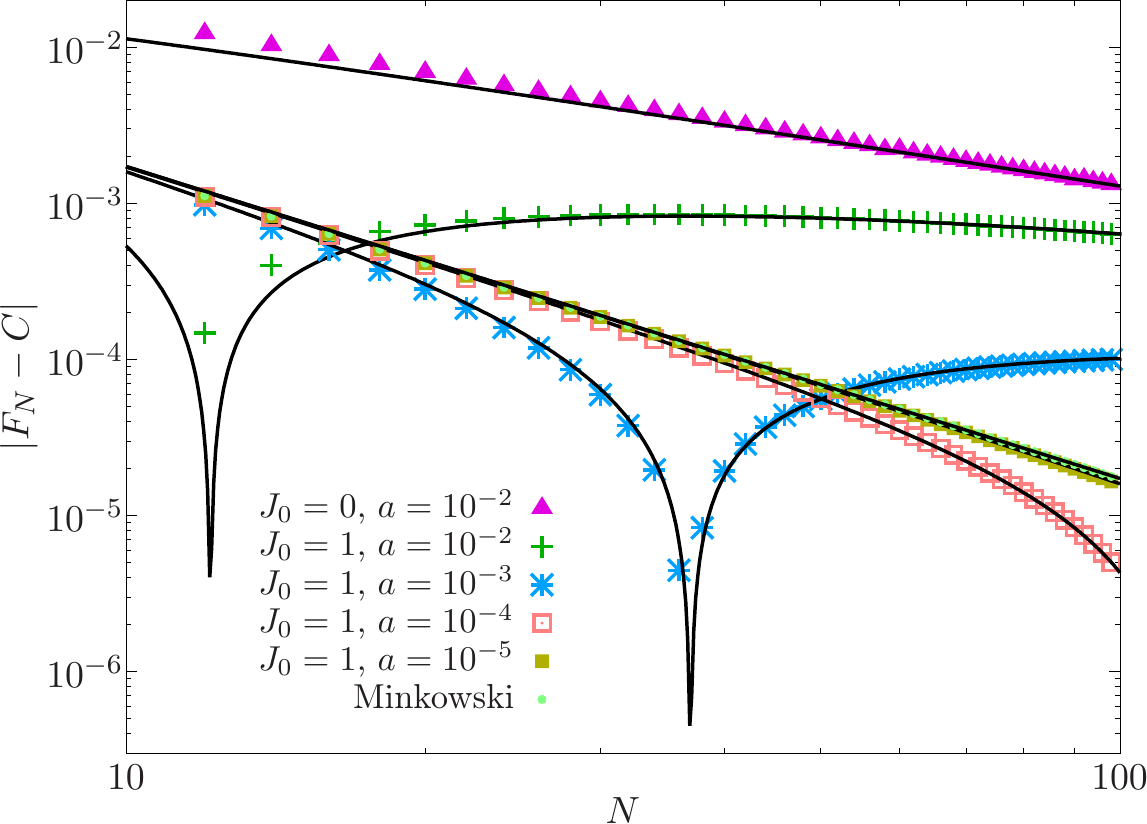}
  \caption{Casimir forces for the spin-1/2 Heisenberg chain with
    Rindler couplings. The black lines correspond to the theoretical
    prediction, given by Eq. \eqref{eq:force_heis}. Compare to the
    inset of Fig. \ref{fig:forces} (top).}
  \label{fig:heisenberg}
\end{figure}

We may wonder whether these results are only valid for free fermions
or if, instead, they can be applied to other CFT. Thus, we have
considered one of the simplest critical interacting systems, the
(inhomogenous) spin-1/2 Heisenberg model in 1D, defined by

\beq
H=-\sum_{i=1}^{L-1} J_i\; \vec S_i \cdot \vec S_{i+1},
\label{eq:heis}
\eeq
Using the Jordan-Wigner transformation we may rewrite it in fermionic
language as

\beq
H=-\sum_{i=1}^{L-1} J_i \(c^\dagger_i c_{i+1}+\text{h.c.}\) + 2 \sum_{i=1}^L J_i\,
n_i n_{i+1},
\label{eq:heis_ferm}
\eeq
where we can see that fermionic particles at nearby sites repel each
other, making it impossible to use free-fermion techniques. Yet, the
GS energy of this Hamiltonian can be accurately obtained using the
density matrix renormalization group (DMRG) algorithm
\cite{White_92,White_93,Details}. The results for the Rindler
couplings, Eq. \eqref{eq:rindler_J} are shown in
Fig. \ref{fig:heisenberg}. The maximal size that we have reached is
lower than in the previous case, $N=100$, because the numerical
computation is more demanding. Yet, the results show that a
straightforward extension of Eq. \eqref{eq:force} predicts the force
values with a remarkable accuracy using $c_0=0.4431$, $c_B=0.2618$ and
$v_F=1.319$, through

\beq
F_N \approx -c_0 - {c_B\over 2} \({J'_N\over J_N}\)
+ {\pi v_F\over 24 N^2}.
\label{eq:force_heis}
\eeq

Fig. \ref{fig:heisenberg} shows $|F_N+c_0|$ in logarithmic scale as a
function of $N$ for different Rindler deformations of the Heisenberg
Hamiltonian, along with the theoretical prediction,
Eq. \eqref{eq:force}. These plots can be compared with the inset of
Fig. \ref{fig:forces} (top).

\section{Conclusions and further work}
\label{sec:conclusions}

We have derived an expression for the ground-state energy of the
discretized version of the Dirac equation in a deformed (1+1)D medium,
which corresponds to the vacuum state in static curved metrics. We can
model a classical particle navigating through the system depressing a
local hopping, and then it can be readily checked that the classical
particle moves approximately in a potential which corresponds to the
classical gravitational potential associated with the metric. The
quantum corrections to this semi-classical result can be obtained by
suitably deforming the predictions of conformal field theory
(CFT). Indeed, we have checked that the finite-size corrections are
dominated by two terms: a boundary term related to the derivative of
the local hopping amplitude at the edge of the system, and a naturally
deformed version of the CFT force, where the central charge is
preserved. The conformal correction can be interpreted in two
complementary ways: either the Fermi velocity is substituted by the
(harmonic) average value of the hopping terms, or the system size is
transformed by its deformed value.

In any case, we should emphasize that the finite-size corrections to
the vacuum energy are, indeed, {\em universal}. Moroever, we have
shown that an observer at a boundary measuring the Casimir forces will
obtain a metric-independent value.

It is relevant to ask whether our results extend to other conformal
field theories, both interacting, such as Heisenberg, or
non-interacting, such as the Ising model in a transverse field. Even
more challenging will be to extend these results to (2+1)D field
theories and to consider non-static metrics, where the dynamical
effects will be relevant, linking them to the dynamical Casimir effect
\cite{Nation.12}. Even if the energy is not defined in those cases, a
force can still be found acting on classical particles. It is also
interesting to consider chains under strong inhomogeneity or
randomness \cite{Ramirez.random.14,Laguna.16,Alba.19,Samos.19b}.

As a natural next step, we intend also to develop protocols in order
to confirm these results in the laboratory employing ultra-cold atoms
in optical lattices, where similar curved-metric problems have been
addressed in the past, such as the measurement of the Unruh
effect \cite{Boada.11,Laguna_unruh.17}. 

\bigskip


\begin{acknowledgments}
  
We thank C. Fernández-González, P. Rodríguez-López, N. Samos Sáenz de
Buruaga and G. Sierra for very useful discussions. Also, we acknowedge
the Spanish government for financial support through grants
PGC2018-094763-B-I00 (SNS) and PID2019-105182GB-I00 (JRL and BMM), and
the Fondo de Garantía Juvenil through contract PEJD-2017-PRE/TIC-4649
(BMM).
\end{acknowledgments}


\appendix

\section{CFT derivation of the Casimir energy in curved backgrounds}
\label{appendix1}

Let us provide a theoretical justification for our deformed extension
of expression \eqref{eq:cft}, given in Eq. \eqref{eq:cft_curved}. The
two first terms are non-universal: $c_0(N-1)\mapsto c_0 S_N$, while
$c_B\mapsto (c_B/2)(J_1+J_N)$ are just a consequence of first-order
perturbation theory. Yet, the finite-size correction term ($c\pi
v_F/24 N$) is universal, i.e. fixed by conformal invariance, and
requires further explanation. In what follows we will assume that the
Fermi velocity (the speed of light) is $v_F=1$.

According to CFT, the variation of the energy-momentum tensor $T$
under a local conformal transformation, $z \rightarrow w(z)$, in flat
spacetime is given by \cite{DiFrancesco}

\beq
T'(w)=\(\dfrac{dw}{dz}\)^{-2}\[ T(z)-\dfrac{c}{12} \{w;z\}\],
\label{eq:T}
\eeq
where $c$ is the central charge of the CFT and $\{w;z\}$ is the
Schwarzian derivative,

\beq
\{w;z\}= \dfrac{d^3w/dz^3}{dw/dz}-
\dfrac{3}{2}\(\dfrac{d^2w/dz^2}{dw/dz}\)^2.
\label{eq:Sch_der}
\eeq

Let us consider a CFT defined on the whole complex plane, with
vanishing energy density $\<T(z)\>\sim 0$. Now, we would like to map it
into a strip of width $L$, using

\beq
z \rightarrow w= \dfrac{L}{\pi} \ln{z}.
\label{eq:conf_transf}
\eeq
This yields a nonzero vacuum energy density on the strip

\beq
\< T_{\text{strip}}(w) \> = - \dfrac{c \pi^2}{24\,L^2}.
\label{eq:Tcyl}
\eeq
Now, the energy density can be evaluated (check Eq. (5.40) of
\cite{DiFrancesco}),

\beq
\<T^{00}\>=\<T_{zz}\>+\<T_{\bar z\bar z}\>=-{1\over\pi}\<T\>={\pi c\over
  24 L^2},
\eeq
which corresponds to the universal term in Eq. \eqref{eq:cft}. Yet,
our $z$ variable is composed of a deformed space variable and time,
$z=\tilde x + it$, so the length appearing in this expression is,
in fact, $\tilde L$, as required.

\medskip

Let us provide an alternative derivation, only valid for infinitesimal
deformations of the metric, $g_{\mu\nu}\mapsto g_{\mu\nu}+\delta
g_{\mu\nu}$. The free energy density of a conformal system, $F$,
varies as

\beq
\delta F = -\dfrac{1}{2} \int d^2x\; \sqrt{g}\; \delta g_{\mu\nu} \<T^{\mu\nu}\>,
\label{Var_F}
\eeq
where $\sqrt{g}=\det\(g_{\mu\nu}\)^{1/2}$ is required by the
invariance of the spacetime integration measure. Let consider the
Minkowski energy density, given by 

\beq
T^{00}= \dfrac{\pi c}{24\, L^2},
\label{T00}
\eeq
and deform the metric, mapping $g_{00}=-1$ to $g_{00}+\delta
g_{00}=-J^2(x)\approx -1-2\delta J(x)$. This leads to a new free energy,

\beq
\delta F = \int d^2x\; \delta J(x) {\pi c\over 24\, L^2},
\eeq
where the integration must be performed on a strip $[0,L]\times
\R$, where the vertical direction is trivial. The total energy is
given by the new free energy per unit length (in the transverse
direction), 

\beq
E = F_L +\delta F_L =
\({1\over L} \int_0^L dx\,(1+\delta J(x))\)  {\pi c\over 24\, L},
\eeq
i.e. the energy gets corrected by a new Fermi velocity, which is equal
to the average value of $J(x)$ in the interval. This is the main
result of Eq. \eqref{eq:deformed}.

Of course, this result is only valid for very small deformations,
$J(x)\approx 1+\delta J(x)$. The full expression \eqref{eq:cft_curved}
can be obtained by integrating it, $F=\int \delta F$. We may
parametrize the change from $g_{00}=-1$ to $g_{00}=-J^2(x)$ in a
continuous way, considering a one-parameter metric family,
$g_{00}(s)=J^2(x,s)$ such that $J^2(x,0)=-1$ and $J^2(x,1)=J^2(x)$, so
that the final energy correction takes the form

\beq
\Delta F = \int_0^1 ds \int dx \sqrt{g(s)}
\(\dfrac{\pi c}{24 L(s)^2}\) {\partial J(x,s)\over \partial s},
\label{eq:F}
\eeq
where $L(s)$ and $\sqrt{g(s)}$ correspond respectively to the
effective length and the volume factor at each stage of the
deformation process.


\section{Casimir force measured by local observer}
\label{appendix2}

Let $E$ be the Casimir energy for the whole system. When it is
measured by a local observer at site $x$ will be given by $E(x)=
E/g_{00}(x)^{1/2}=E/J(x)$, following
Eq. \eqref{eq:energy_observer}. Let us remember that the energy is not
a scalar, but a vector pointing along the time axis: $(E(x),0)$. The
force is defined as the spatial component of the covariant derivative
of the energy,

\beq
F(x)=-D_x E(x),
\eeq
where the covariant derivative of a vector is defined as

\beq
D_\mu V^\alpha = \partial_\mu V^\alpha + \Gamma^\alpha_{\mu\nu} V^\nu,
\label{eq:covariant}
\eeq
where the $\Gamma^\alpha_{\mu\nu}$ are the Christoffel symbols, given
by

\beq
\Gamma^\alpha_{\mu\nu}={1\over 2}g^{\alpha\beta} \(
g_{\beta\mu,\nu}+g_{\beta\nu,\mu}-g_{\mu\nu,\beta}\).
\label{eq:christoffel}
\eeq
for the metric connection. In the case of an optical metric,
Eq. \eqref{eq:metric}, the only relevant Christoffel symbol is

\beq
\Gamma^0_{10}={J'(x)\over J(x)}.
\eeq

Thus, we can find the force

\beq
F(x)=-{\partial_x E(x)\over J(x)} + {J'(x)\over J(x)} E(x) -
{J'(x)\over J(x)} E(x) = -{\partial_x E(x)\over J(x)}.
\eeq
And from this equation we can find a possible definition of the
Casimir force felt by a local observer at the boundary,

\beq
F_N \approx -{E_N-E_{N-1}\over J_N \Delta x},
\eeq
where we set $\Delta x=1$, since it is arbitrary. Yet, the strong
parity oscillations suggest that a better alternative is to take the
discrete derivative over two lattice spacings,

\beq
F_N \equiv -{E_N-E_{N-2} \over J_{N-2}+J_{N-1}}.
\eeq


\end{document}